\documentclass[sigconf]{acmart}
\usepackage{xspace}
\usepackage[capitalize,noabbrev,nameinlink]{cleveref}
\usepackage{hyperref}  
\usepackage{diagbox}
\usepackage{multirow}
\usepackage{enumitem}
\usepackage{balance}
\DeclareMathOperator*{\argmax}{arg\,max}

\renewcommand\footnotetextcopyrightpermission[1]{}

\begin{document}

\title{Semantic IDs for Recommender Systems at Snapchat: \\ Use Cases, Technical Challenges, and Design Choices}

\author{Clark Mingxuan Ju$^*$, Tong Zhao$^*$, Leonardo Neves, Liam Collins, Bhuvesh Kumar, Jiwen Ren, Lili Zhang, Wenfeng Zhuo, Vincent Zhang, Xiao Bai, Jinchao Li, Karthik Iyer, Zihao Fan, Yilun Xu, Yiwen Chen, Peicheng Yu, Manish Malik, Neil Shah}
\email{{mju, tong, nshah}@snap.com}
\affiliation{%
  \institution{Snap Inc.}
  \city{Santa Monica}
  \state{CA}
  \country{USA}
}
\renewcommand{\shortauthors}{Clark Mingxuan Ju and Tong Zhao et al.}
\renewcommand{\shorttitle}{Semantic IDs for Recommender Systems at Snapchat: Use Cases, Technical Challenges, and Design Choices}

\begin{abstract}
  Effective item identifiers (IDs) are an important component for recommender systems (RecSys) in practice, and are commonly adopted in many use cases such as retrieval and ranking.
  IDs can encode collaborative filtering signals within training data, such that RecSys models can extrapolate during the inference and personalize the prediction based on users' behavioral histories. 
  Recently, Semantic IDs (SIDs) have become a trending paradigm for RecSys. 
  In comparison to the conventional atomic ID, an SID is an ordered list of codes, derived from tokenizers such as residual quantization, applied to semantic representations commonly extracted from foundation models or collaborative signals.  
  SIDs have drastically smaller cardinality than the atomic counterpart, and induce semantic clustering in the ID space.  At Snapchat, we apply SIDs as auxiliary features for ranking models, and also explore SIDs as additional retrieval sources in different ML applications.
  In this paper, we discuss practical technical challenges we encountered while applying SIDs, experiments we have conducted, and design choices we have iterated to mitigate these challenges.
  Backed by promising offline results on both internal data and academic benchmarks as well as online A/B studies, SID variants have been launched in multiple production models with positive metrics impact. 
\end{abstract}
\maketitle
\begingroup
\renewcommand\thefootnote{}\footnote{$^*$Equal contributions.}
\renewcommand\thefootnote{}\footnote{Code is publicly available at \url{https://github.com/snap-research/GRID}.}
\addtocounter{footnote}{-1}
\endgroup
\vspace{-0.9cm}
\section{Introduction}
The success of modern social media ecosystems hinges on recommendation systems (RecSys)~\citep{pal2020pinnersage,gomez2015netflix,van2013deep,schafer1999recommender}, which are pivotal in personalizing users' interactions across various applications, ranging from short-form video feeds~\citep{ju2025learning,deng2025onerec,singh2024better,he2025plum,ju2025generative,ju2025revisiting} and friend recommendation~\citep{shi2023embedding,tang2022friend,kolodner2024robust} and e-commerce platforms~\citep{chen2025onesearch,hou2024bridging,wang2021dcn,schafer1999recommender}.
RecSys models are usually trained with user historical interaction data (e.g. videos watched, merchandise purchased, friends made, etc.) so that they can predict user's future intents to facilitate better personalization~\citep{rendle2009bpr,kang2018self}. 
Albeit different RecSys model architectures, one important ingredient that broadly exists in these models is the use of sparse \emph{identifiers} (IDs), due to their strong capabilities in encoding the historical interactions among users and items~\citep{singh2024better,cheng2016wide,yuan2023go,wu2025graphhash}.
Specifically, each user or item is assigned with an ID which serves as a unique discrete token that is mapped to a learnable dense embedding vector, compressing sparse interactions into continuous latent embeddings capturing preferences or properties.

While effective, the reliance on per-user/item atomic IDs introduces significant challenges. 
First, the sheer magnitude of users and items leads to an explosion in parameter size, as massive embedding tables are required to store distinct vectors for every entity~\citep{zhang2020model,rendle2009bpr,wu2025graphhash}. 
Second, ID-based approaches lack inductive support, leaving the model unable to generate representations for new users or items (the cold-start problem) that lack a pre-existing entry in the embedding table~\citep{shiao2025improving,zhao2023embedding}. 
Third, these models struggle with generalization due to the long-tail distribution of data; popular IDs receive the vast majority of gradient updates, resulting in robust embeddings, while the representations for tail entities remain under-trained and sub-optimal due to sparse supervision~\citep{singh2024better,ju2024does}.

To address these challenges, \citet{rajput2023recommender} and \citet{singh2024better} proposed replacing conventional atomic IDs with Semantic IDs (SIDs) -- a strategy subsequently validated by practitioners for its promising recommendation performance~\citep{ju2025generative,deng2025onerec,yang2024unifying,liang2026generative,lee2026sequential,chen2024enhancing}. 
Each SID consists of an ordered sequence of discrete tokens, where the cardinality at each index is significantly lower than that of traditional user or item IDs. 
This paradigm usually employs a semantic or modality encoder (e.g., LLMs or VLM) paired with a residual quantization tokenizer -- such as RQ-VAE~\citep{lee2022autoregressive}, VQ-VAE~\citep{esser2021taming}, or Residual K-means~\citep{deng2025onerec} -- to transform raw modality features (e.g. image, text, audio) into discrete SIDs. 
Intuitively, SIDs function as a semantic hashing mechanism that map semantically similar items or users to discrete codes sharing common prefixes.
% SIDs provide an effective mechanism to leverage both the semantic knowledge encoded in pre-trained foundation models and the collaborative signals derived from user-item interaction histories; specifically, the overlap in SIDs reflects semantic similarity, while training supervision enables the model to capture collaborative dynamics.

Given the potential of SIDs, we conducted extensive experiments at Snapchat~\citep{ju2025generative}, exploring and identifying integration opportunities as auxiliary features into ranking and retrieval stacks to improve recommendation performance across diverse surfaces (e.g., ads, content, and growth).  
This approach enables RecSys models to incorporate rich semantic signals without incurring the prohibitive computational costs associated with high-dimensional dense embeddings. 
Furthermore, it promotes robust model generalization by mitigating cold-start challenges and maximizing parameter utilization, as the shared hierarchical structure of SIDs ensures that token embeddings receive frequent gradient updates across the entire item catalog.
Beyond feature representation, we expanded our exploration to generative retrieval (GR)~\citep{rajput2023recommender,ju2025generative,liu2025understanding,zhu2025beyond}, using SIDs as the primary retrieval target in our content recommendation. 
Specifically, we replaced atomic IDs in user behavior sequences with SIDs and trained a sequential recommender to generate the SID token sequences of future items based on historical interactions.

However, utilizing SIDs at an industrial scale presented technical hurdles.
First, we encountered severe codebook collapse during the training of the RQ-VAE tokenizer, where the model utilized only a fraction of the available codebook, severely limiting the semantic expressiveness of the generated IDs.
Second, utilizing SIDs as a retrieval source introduced complex SID-to-Item resolution challenges, particularly in handling collisions and ensuring valid decoding.
In this work, we conduct extensive experiments on internal data at Snap and detail the specific architectural modifications we implemented in RQ-VAE to effectively mitigate codebook collapse.
Furthermore, we present a comprehensive analysis of the strategies and heuristics we developed for SID-to-Item resolution. 
Our contributions are summarized as follows:
\begin{itemize}[leftmargin=*,topsep=0pt]
\item \textbf{Industrial-scale Evaluation}: We present a large-scale application of SIDs at Snapchat, demonstrating their efficacy across diverse surfaces (e.g., ads, content, growth, and search), reporting significant gains in both offline and offline metrics.
\item \textbf{Technical Improvements}: We propose novel architectural modifications to the RQ-VAE tokenizer that effectively mitigate codebook collapse. Additionally, we provide a detailed analysis of SID-to-Item resolution strategies, offering practical heuristics to ensure valid and accurate decoding during retrieval.
\item \textbf{Open-source Contribution}: To facilitate reproducibility and future research, we open-source our optimized model architecture and training implementations for the broader community.
\end{itemize}
\vspace{-0.2cm}
\section{Preliminaries}
This section introduces the notation for SIDs and details the formulation of RQ-VAE~\citep{lee2022autoregressive}, a widely adopted tokenizer for SID generation~\citep{rajput2023recommender,liang2026generative,he2025plum,luo2025qarm}. 
We focus on RQ-VAE due to its differentiable nature, which seamlessly supports customized extensions like multi-embedding fusion, though we recognize that non-differentiable approaches like RQ-Kmeans~\citep{deng2025onerec,ju2025generative} are also highly effective.
Given an item or user $i$ with semantic features $x_i$, we use a modality encoder $\phi(\cdot): x \rightarrow \mathcal{R}^d$ (e.g., LLM, VLM, etc) to transform the semantic feature into the hidden representation $\mathbf{h}_i \in \mathcal{R}^d$, where $d$ refers to the output dimension of $\phi(\cdot)$. 
RQ-VAE can be denoted as $\text{RQ-VAE}(\cdot; \mathbf{\theta}): \mathcal{R}^d \rightarrow \{0, 1, \dots, K-1\}^L$, where $K$ refers to the number of IDs in each layer, $L$ refers to the total number of codebooks, and $\theta$ refers to model parameters including an encoder, a decoder, and $L$ codebooks. 
The encoder $\text{Enc}(\cdot): \mathcal{R}^d \rightarrow \mathcal{R}^n$ transforms $\mathbf{h}_i$ into the intermediate embedding $\mathbf{h}_i^0$ (i.e., 0-th residual before the residual quantization), where $n$ refers to the hidden dimension of RQ-VAE. 
Then, $\mathbf{h}_i^0$ is used to calculate the first SID code $\text{sid}_0$ with the first codebook $\mathbf{C}_0 \in \mathcal{R}^{K \times n}$, formulated as:
\vspace{-0.05in}
\begin{equation}
    \text{sid}_0 = \argmax_{c \in \{0, 1, \dots, K-1\}} \big|\big|\mathbf{h}_i^0 \cdot \mathbf{C}_0[c]\big|\big|_{\text{F}} \text{ ,with } \mathbf{h}_i^0=\text{Enc}\big(\phi(x_i)\big),
    \label{eq:sid}
    \vspace{-0.05in}
\end{equation}
where $||\cdot||_\text{F}$ is Frobenius norm and $[\cdot]$ is the indexing operation. 
Following \cref{eq:sid}, the later SID codes $\text{sid}_l$ where $l \geq 1$, they can be acquired by residual quantization as: 
\vspace{-0.05in}
\begin{equation}
    \text{sid}_l = \argmax_{c \in \{0, 1, \dots, K-1\}} \big|\big|\mathbf{h}_i^l \cdot \mathbf{C}_l[c]\big|\big|_{\text{F}}, % \text{ ,with } \mathbf{h}_i^l= \mathbf{h}_i^{l-1} - \mathbf{C}_{l-1}[\text{sid}_{l-1}].
    \vspace{-0.03in}
\end{equation}
where $\mathbf{h}_i^l= \mathbf{h}_i^{l-1} - \mathbf{C}_{l-1}[\text{sid}_{l-1}]$.
Once all SID codes are derived, their corresponding centroids are aggregated and fed into the decoder for reconstruction, as the following:
\vspace{-0.05in}
\begin{equation}
    \hat{\mathbf{h}_i} = \text{Dec}\Big(\sum_{l \in \{0,\dots,L-1\}} \mathbf{C}_l[\text{sid}_l]\Big),
    \label{eq:dec}
    \vspace{-0.05in}
\end{equation}
where $\text{Dec}(\cdot): \mathcal{R}^n \rightarrow \mathcal{R}^d$ is the decoder and $\hat{\mathbf{h}_i} \in \mathcal{R}^d$ is the reconstructed representation. 
RQ-VAE is supervised jointly by a reconstruction loss and a commitment loss~\citep{lee2022autoregressive}; the former trains the model to faithfully reconstruct the input data , and the latter prevents codebooks from growing boundlessly and causing collapse. 
% \ns{Sec 3 reads a bit awkwardly because we talk about SID-specific codebook challenge, and then a resolution challenge which only is encountered during Generative Retrieval.  This part is not mentioned in Sec 2, which doesn't talk about Gen Retrieval at all but ust talks about RQ-VAE.}
\section{Technical Challenges in Applying SIDs}
While SID presents a promising paradigm for RecSys, application in practice is non-trivial and entails multiple key challenges. 
Specifically, these hurdles emerge both during the initial training of the RQ-VAE tokenizer to create robust SIDs, and during the subsequent application of these SIDs within downstream GR frameworks.

\noindent \textbf{Challenge 1: Codebook Collapse}. The first significant challenge we encountered occurred during the training of the RQ-VAE tokenizer, manifesting as severe codebook collapse.  
In this scenario, the model utilized only a fraction of the available codebook, assigning items to only a few codes. 
This under-utilization severely limits the semantic expressiveness of the generated IDs.
Following the quantization process formulated in \cref{eq:sid}, the intermediate hidden representations $h_{i}^{l}$ are expected to map across a diverse set of centroids within the codebook $C_{l}$. 
However, due to the imbalanced training dynamics often observed in standard RQ-VAE implementations~\citep{liang2026generative,zhu2024scaling}, a small subset of popular centroids receive the vast majority of gradient updates. 
Meanwhile, remaining centroids are effectively ``dead''. 
Because SIDs act as a semantic hashing mechanism, an impoverished codebook limits the granularity of the hashing, bottlenecking the model's ability to accurately differentiate between fine-grained item features.

We identified two design choices to mitigate collapse.  Firstly, we \emph{backpropagate through the entire codebook} using the straight-through estimator (STE) \cite{bengio2013estimating}. 
In the standard RQ-VAE formulation, codebook optimization is inherently discrete due to the $\argmax$ operation in \cref{eq:sid}. 
As a result, during backprop, gradient updates are exclusively applied to the specific codebook rows (centroids) that are actively selected by the input representations. 
This sparse update mechanism causes training to rely heavily on the initial codebook state; poor initialization frequently leads to many unselected, ``dead'' centroids which never receive gradient updates.
To address this, we apply STE to approximate the gradients through quantization. Specifically, \cref{eq:dec} is modified as the following:
\vspace{-0.05in}
\begin{equation*}
    \hat{\mathbf{h}_i} = \text{Dec}\Big(\sum_{l \in \{0,\dots,L-1\}} \mathbf{C}_l[\text{sid}_l] + \text{sim}(\mathbf{h}_i^l, \mathbf{C}_l)\cdot\mathbf{C}_l - \text{sg}[\text{sim}(\mathbf{h}_i^l, \mathbf{C}_l)\cdot\mathbf{C}_l]\Big),
    \vspace{-0.05in}
\end{equation*}
where $\text{sim}(\mathbf{h}_i^l, \mathbf{C}_l) \in \mathcal{R}^K$ refers to the cosine similarities between the residual at layer $l$ and the entire codebook at layer $l$, and $\text{sg}(\cdot)$ refers to the stop-gradient operation, which prevents encompassed variables from contributing to the backprop graph. 
By utilizing STE, the loss can backprop through the assignment mechanism, effectively involving the entire codebook in updates. This promotes continuous update of all codes and promotes more uniform utilization of available codes, significantly reducing collapse risk. 
% \ns{normal STE would be like using argmax in forward and softmax gradients in backward.  looks like we actually introduced a new path or term with this $sim(h_i^l, C_l)\cdot C_l$ thing. feels a bit unlike normal STE but i agree it resembles.  not sure if we should describe differently to be more correct.}

The second choice we adopted to mitigate collapse is to learn SIDs \emph{conditioned on multiple embedding sources}.  
In industrial RecSys, items rarely exist as single-modality entities; rather, they are characterized by rich, heterogeneous features such as textual descriptions, visual content, categorical metadata, etc. 
We observed relying on a single embedding source (e.g. text) yields a more homogeneous input distribution. 
This lack of variance exacerbates collapse, as the RQ-VAE requires fewer centroids to adequately encode and reconstruct the input.
Instead, we rely on an embedding input fusion approach -- specifically, we aggregate disparate pre-trained embeddings into a unified continuous representation before feeding them into the tokenizer. 
Using a summation aggregation, the encoding and decoding processes for an item $i$ are formulated as:
\begin{equation*}
    \mathbf{h}_i = \sum_{m \in M} \text{Enc}_m(x_m) \text{ and } \hat{\mathbf{h}_{i,m}} = \text{Dec}_m\Big(\sum_{l \in \{0,\dots,L-1\}} \mathbf{C}_l[\text{sid}_l]\Big).
\end{equation*}
By integrating multi-modal signals, the resulting input space exhibits significantly higher variance and topological complexity. 
This richer input distribution forces the quantization process to span a wider manifold to minimize reconstruction loss. 

\noindent \textbf{Challenge 2: SID-to-Item Resolution}. A second hurdle in practical SID use is the challenge of grounding in the context of GR, which amounts to directly resolving a single item from a SID. 
Since SIDs map semantically similar items or users to discrete codes with common prefixes, where the cardinality of each codebook layer is vastly smaller than atomic IDs, similar items often map to the exact same discrete token sequence (collision). Resolving these collisions and determining which specific item(s) within a shared code should be retrieved or ranked requires additional disambiguation.

We found two helpful choices in managing this resolution problem.  Firstly, we adopt a \emph{heuristic-based intra-code disambiguation}.
Specifically, if a generated SID has multiple associated items, we employ a secondary ranking mechanism based on domain-specific heuristics. 
Instead of treating all items mapping to the SID as equally relevant, we resolve an item by incorporating additional item-level metadata during the retrieval phase.  For example, in a video  retrieval setting, we may leverage historical relevance metrics or temporal features, e.g. cumulative view time or content freshness, to prioritize higher-quality or more business-metric-relevant items.  This two-stage approach allows the generative model to identify the correct semantic category (SID) while relying on lightweight heuristics to pinpoint the most relevant items.

Secondly, we found it useful to prioritize \emph{retrieval depth over breadth}.  
During inference, an GR model outputs a set of candidate SIDs via beam search or sampling. 
This introduces a practical trade-off given a fixed retrieval budget: fetching a few items across many different candidate SIDs (breadth) versus fetching many items from a few top-ranked SIDs (depth).
Through empirical evaluation, we discovered that maximizing item yield from the top-ranked SIDs significantly outperforms distributing the item budget across a larger number of lower-ranked SIDs. This indicates that the model possesses high confidence and accuracy in its top-ranked predictions. 
\section{Experiments}
To comprehensively evaluate the efficacy of SIDs for RecSys, we conduct a series of offline and online experiments.

\subsection{SID as Auxiliary Features}
In this set of experiments, we investigate the performance of SIDs by incorporating them as auxiliary categorical features within standard ranking and retrieval architectures.
% (e.g. two-tower models) \ns{these archs might actually be like DCN or MMoE or PLE archs which are not really two-tower models}. 
The objective is to determine whether the hierarchical, multimodal semantic priors captured by SIDs can improve model generalization.
Specifically, we applied SIDs to multiple RecSys models at Snapchat, as the following:
\begin{itemize}[leftmargin=*,topsep=0pt]
\item \textbf{Ads Ranking}. 
Ads are typically accompanied by textual metadata, such as descriptive titles, brand names, and category labels. 
We utilize the Qwen-Embedding model~\citep{yang2025qwen3} to encode these text into embeddings, which are subsequently converted into SIDs. 
This approach allows the ranking model to leverage rich semantic information without the large overhead of logging and training with raw, high-dimensional embeddings.
As shown in \cref{tab:sid_feature}, adding SIDs improved the Swipe Up and Landing Page View metrics (goal-based bidding) by 0.028\% and 0.035\%, respectively.
\item \textbf{Dynamic Product Ads (DPA) Ranking}.
DPA models encounter a special challenge of rapid item ID churn due to constantly updating advertiser catalogs. 
Following a similar pipeline, we use an LLM to encode product metadata into embeddings, which are then converted into SIDs. 
Demonstrated in \cref{tab:sid_feature}, SIDs provided even more substantial gains in this high-churn environment, driving a +0.67\% improvement in Add to Cart predictions and a +0.24\% average improvement across all heads.
\item \textbf{Friending and Search Ranking}.
In the context of user friend recommendation and search, we deploy GraphHash~\cite{wu2025graphhash} as a unique form of SID learned directly from graph semantics, where the underlying network topology is defined by friending user interactions. By hierarchically clustering users into structural communities, GraphHash compresses over 900M raw user IDs into a condensed set of SIDs. As auxiliary features, GraphHash SIDs enable the model to extrapolate collaborative filtering signals more effectively. Consequently, adding GraphHash SIDs has yielded strong online improvements, as \cref{tab:sid_feature_ab} shows, driving meaningful gains in product metrics (e.g., Creator subscribe, reciprocation rate, etc) and materially reducing negative friending actions (e.g., friending blocks/deletes).
% \ns{shall we actually use metric names similar to what we do for Ads and Content?}.
\end{itemize}
\begin{table}[t]
    \centering
    \vspace{-0.1in}
    \caption{Offline improvements to AUCs when we SIDs are incorporated as auxiliary input features to different rankers, where a 0.01\% gain is considered significant. }
    \vspace{-0.15in}
    \small
    \begin{tabular}{l|c}
    \toprule 
    Metrics &  Improvements\\ 
    \midrule
    \multicolumn{2}{c}{\textit{Ads Ranking}} \\ 
    \midrule
    Swipe up (goal-based bidding) & +0.028\% \\ % ±0.09\% \\
    Landing page view (goal-based bidding) & +0.035\% \\ % ±0.04\% \\
    \midrule
    \multicolumn{2}{c}{\textit{Dynamic Product Ads Ranking}} \\ 
    \midrule
    Add to cart & +0.67\% \\
    Average improv. cross all heads & +0.24\% \\
    \bottomrule
    \end{tabular}
    \label{tab:sid_feature}
    \vspace{-0.05in}
\end{table}
\begin{table}[t]
    \centering
    \vspace{-0.05in}
    \caption{Online improvements of launching GraphHash SID as auxiliary feature in different rankers.}
    \small
    \vspace{-0.1in}
    \begin{tabular}{l|cc}
    \toprule 
    Ranker Models & metrics &  online improv.\\ 
    \midrule
    Find Friends & relevance & +1.77\%, +4.90\%\\
    Friending Registration & neg. action & -5.13\%, -3.17\% \\
    Display Name Search & feature use & +1.55\%, +12.43\% \\
    % \midrule
    \bottomrule
    \end{tabular}
    \label{tab:sid_feature_ab}
    \vspace{-0.2in}
\end{table}

\subsection{SIDs in Generative Retrieval}
We also explore their efficacy as the primary target space for GR within our content recommendation stack. 
In this paradigm, we replace conventional atomic IDs in user behavior sequences with SIDs and train a sequential recommender to autoregressively generate the SID sequences~\citep{rajput2023recommender,ju2025generative}. 
We evaluate the performance of this GR framework in our content retrieval stack and discuss the practical design choices including heuristic-based intra-bucket disambiguation and depth-prioritized fetching required to enable collision resolution and higher-quality recommendations.
Detailed model architecture is described in our previous work~\citep{ju2025generative}.

As shown in \cref{tab:sid_retrieval_offline}, providing the sequential recommender with longer user behavior sequences substantially boosts retrieval accuracy (e.g., increasing the sequence length from a 120 baseline to 480 yielded massive gains of +31.5\% for R@5 and +26.5\% for N@5).
Balancing infrastructure constraints, we stop at 480 and use this variant as the launch candidate. 
Validated by promising offline evaluation, we conducted A/B studies, as shown in \cref{tab:sid_retrieval_ab}.
Online experiments in short-form video recommendations highlight the critical importance of intra-bucket disambiguation strategies. 
While random mapping from the top SIDs yielded marginal gains (e.g., +0.13\% for video views), utilizing an relevance-guided mapping strategy from the top 10 SIDs drove substantial increases in high-intent user actions. 
Specifically, this heuristic-based approach resulted in a +0.57\% increase in views, a +2.54\% increase in sends, a +3.55\% increase in re-posts, and a remarkable +4.39\% increase in short-form video shares in an online experiment.
\begin{table}[t]
    \centering
    \caption{Offline ranking-based metrics (R/N refer to recall/NDCG) when SIDs are used as direct retrieval 
 in generative retrieval for short-form video recommendation.  }
    \small
    \vspace{-0.1in}
    \begin{tabular}{l|c|c|c|c}
    \toprule 
    Model Variant & R@5 & R@10 & N@5 & N@10\\ 
    \midrule
    Seq. Length = 120 &  \multicolumn{4}{c}{Baseline} \\ % ±0.09\% \\
    \midrule
    Seq. Length = 240 & +13.7\% & +8.3\% & +8.2\% & +6.5\% \\ 
    Seq. Length = 480 & +31.5\% & +16.7\% & +26.5\% & +22.2\% \\ 
    \bottomrule
    \end{tabular}
    \vspace{-0.1in}
    \label{tab:sid_retrieval_offline}
\end{table}
\begin{table}[t]
    \centering
    % \vspace{-0.05in}
    \caption{Online improvements in A/B testing for short-form video recommendation when SIDs are used as direct retrieval sources with sequence length equal of 480.}
    \small
    \vspace{-0.1in}
    \begin{tabular}{l|c}
    \toprule 
    Metrics &  Improvements (\%)\\ 
    \midrule
    \multicolumn{2}{c}{\textit{Random Item Mapping (10 per SID) from Top 100 SIDs}} \\ 
    \midrule
    % Short-form video view & Neutral\\
    Short-form video view & Neutral\\
    \midrule
    \multicolumn{2}{c}{\textit{Random Item Mapping (100 per SID) from Top 10 SIDs}} \\ 
    \midrule
    Short-form video view & +0.13\% \\
    \midrule
    \multicolumn{2}{c}{\textit{Relevance-guided Mapping (100 per SID) from Top 10 SIDs}} \\ 
    \midrule
    Short-form video view & +0.57\% \\
    Short-form video send & +2.54\% \\ 
    Short-form video share & +4.39\% \\ 
    Short-form video re-post & +3.55\% \\ 
    \bottomrule
    \end{tabular}
    \label{tab:sid_retrieval_ab}
    \vspace{-0.1in}
\end{table}
\begin{table}[t]
    \centering
    \caption{Uniqueness of SIDs as well as their GR performance on Amazon beauty dataset. A deduplicate token is appended to ensure fair comparison across codebook shapes~\citep{rajput2023recommender}.}
    \small
    \vspace{-0.1in}
    \begin{tabular}{l|c|c}
    \toprule 
    Codebook Shape & Uniqueness & Recall@10 \\ 
    \midrule
    1024 $\times$ 1024 $\times$ 1024 & 92.95\% & 6.1\\ 
    512 $\times$ 512 $\times$ 512 & 91.79\% & 6.2\\ 
    256 $\times$ 256 $\times$ 256 & 81.65\% & 6.1\\ 
    128 $\times$ 128 $\times$ 128 & 70.58\% & 6.0\\ 
    64 $\times$ 64 $\times$ 64 & 65.40\% & 5.8\\ 
    \bottomrule
    \end{tabular}
    \label{tab:sid_uniqueness_amazon}
    \vspace{-0.15in}
\end{table}
\begin{table}[t]
    \centering
    \caption{Uniqueness of SIDs on internal data when SIDs are generated with different treatments.}
    \small
    \vspace{-0.1in}
    \begin{tabular}{l|c}
    \toprule 
    Treatment & Uniqueness Improvement\\ 
    \midrule
    Clip embedding with RQ-VAE & Baseline \\ 
    \; + RQ-VAE with STE & + 83.4\% \\ 
    \;\;\;\; + Multi-modal embedding & + 23.7\%\\ 
    \;\;\;\; + Audio embedding & + 11.2\%\\ 
    \;\;\;\; + Transcript embedding & + 4.7\%\\ 
    \bottomrule
    \end{tabular}
    \label{tab:sid_uniqueness_mm}
\end{table}

\subsection{SID Quality Evaluation: Is Uniqueness the Golden Standard?}
% \ns{one weird part in this paper is that we talk about how uniqueness is not a gold standard.  but then we show in Table 6 and discuss in sec 3 that we make improvements to avoid codebook collapse and promote uniqueness.  I think a reader may feel we are sabotaging ourselves by saying "uniqueness is not a great metric to measure" and then reporting "uniqueness improvement by embedding fusion" in Table 6.  Also i'm not sure readers put together that uniqueness is the opposite of collapse basically.}
During the iterative development of SIDs, a significant challenge lies in identifying a reliable, lightweight proxy to evaluate SID quality prior to executing computationally expensive offline training and online A/B testing. 
In existing literature, \textit{uniqueness} -- typically defined as the ratio of unique used SIDs to the total number of items, which inversely reflects the collision rate -- is widely adopted as the \textit{de facto} golden standard for SID evaluation.
Uniqueness inversely reflects the collision rate; therefore, a collapsed codebook inherently suffers from near-zero uniqueness. 
While our interventions are necessary to rescue the model from this collapsed state by driving uniqueness into a healthy range, we ask the question: 
\begin{center}
    \textit{Is maximizing uniqueness indefinitely always beneficial?}
\end{center}
To investigate this assumption, we analyzed the correlation between SID uniqueness and GR performance on the Amazon Beauty dataset~\citep{geng2022recommendation}. 
Our empirical findings reveal a nuanced, non-linear relationship rather than a strict monotonic correlation. 
As shown in \cref{tab:sid_uniqueness_amazon}, when SID uniqueness is very low, there is indeed a strong positive correlation -- resolving severe item collisions (tending towards collapse) directly aids the autoregressive model in accurately distinguishing target items. 
However, once uniqueness surpasses a certain threshold (empirically 70\% for this experiment), this correlation plateaus. 
Beyond this saturation point, aggressively chasing higher uniqueness yields negligible additional performance boosts.
These results suggest a necessary re-evaluation of how we measure SID quality.
\emph{Uniqueness should not be evaluated as a gold standard, but rather as a foundational sanity check against collapse and baseline item distinguishability.} 
Developing a highly predictive, offline metric that accurately quantifies the optimal balance between semantic richness and item distinctiveness for downstream applications remains an open question.
\section{Conclusion}
In this work, we detailed the large-scale deployment of SIDs across Snapchat's recommendation systems, demonstrating their success as both auxiliary features and primary retrieval sources. 
To overcome critical technical hurdles, we introduced architectural improvements to the RQ-VAE tokenizer -- such as STE optimization and multi-modal embedding fusion -- to the mitigate codebook collapse.
Additionally, we addressed SID-to-item collisions during inference using heuristic-based intra-bucket disambiguation. 
Supported by extensive offline and online A/B testing, these design choices drove significant improvements in core business metrics. 
Finally, our evaluation revealed that SID uniqueness should not be viewed as the definitive golden standard for quality, underscoring the need for future research into better offline metrics that better balance semantic richness with item distinctiveness.
We open-sourced our optimized model architectures, aiming to foster a continued innovation within the broader recommender systems community.
\newpage
\section{Presenter Bio}
Tong Zhao is a research scientist at Snap Inc.
% , helping building products that improve the user's experiences with Snapchat app leveraging cutting-edge techniques including but not limited to graph machine learning, sequential modeling, and natural language processing. 
He obtained his Ph.D. degree in Computer Science and Engineering at University of Notre Dame and published at top-tier conferences in machine learning, artificial intelligence, data mining, and natural language processing.
\vspace{-0.2in}
\bibliographystyle{ACM-Reference-Format}
\bibliography{acmart}

% \appendix
\end{document}